\documentstyle[prl,aps,preprint,epsf]{revtex}

\begin{document}
\newcommand{\be}{\begin{equation}}
\newcommand{\ee}{\end{equation}}
\newcommand{\bq}{\begin{eqnarray}}
\newcommand{\eq}{\end{eqnarray}}
\newcommand{\Sc}{Schr\"odinger\,\,}
\newcommand{\Sp}{\,\,\,\,\,\,\,\,\,\,\,\,\,}
\newcommand{\no}{\nonumber\\}
\newcommand{\tr}{\text{tr}}
\newcommand{\p}{\partial}
\newcommand{\la}{\lambda}
\newcommand{\La}{\Lambda}
\newcommand{\G}{{\cal G}}
\newcommand{\D}{{\cal D}}
\newcommand{\E}{{\cal E}}
\newcommand{\W}{{\bf W}}
\newcommand{\de}{\delta}
\newcommand{\al}{\alpha}
\newcommand{\bi}{\beta}
\newcommand{\ga}{\gamma}
\newcommand{\ep}{\epsilon}
\newcommand{\vep}{\varepsilon}
\newcommand{\th}{\theta}
\newcommand{\om}{\omega}
\newcommand{\J}{{\cal J}}
\newcommand{\pr}{\prime}
\newcommand{\ka}{\kappa}
\newcommand{\TH}{\mbox{\boldmath${\theta}$}}
\newcommand{\DE}{\mbox{\boldmath${\delta}$}}

\setcounter{page}{0}
\def\footnoterule{\kern-3pt \hrule width\hsize \kern3pt}
%\tighten
\title{
Interactions in Abelian and Yang-Mills Theories \footnote{This work is supported in part from the Ministry of Science and Education.}
}
\author{Jiannis Pachos
\footnote{Email address: {\tt ipachos@atlas.uoa.gr}}
}

\address{Department of Physics, University of Ljubljana\\
\centerline{\rm and}
J. Stefan Institute\\
Ljubljana 1000, Slovenia \\
{~}}

\date{IJS-TP-98/25,~~October 1998}

\maketitle

\thispagestyle{empty}

\begin{abstract}

There is a natural way to study the long distance interactions of gauge theories in the electric (momentum) representation. Here, the main ideas are presented for the Abelian and Yang-Mills gauge theories emphasizing on the structure and the advantages of this approach.

\end{abstract}

\vspace*{\fill}

\newpage

\section{Introduction}

For simple $1+1$ dimensional theories the information provided from the existing symmetries is enough to determine the dynamics of the theory (wave functional, correlators etc.). Such examples are the conformal theories \cite{conformal} as well as the pure gravitation \cite{Jackiw10,Jackiw} and Yang-Mills \cite{Migdal,Jiannis}, which in $1+1$ dimensions are easily solved just by considering the constraints the (conformal, general coordinate or gauge) symmetries impose on the wave functional. In these low dimensional field theories, except in the case we incorporate additional fields, the Hamiltonian is trivial (e.g. has only kinetic term) and it is not difficult to find its eigenvalues. Hence, the information of the constraints applied on the Hamiltonian through the properties they impose on the wave functional is enough to determine completely the dynamics. 

In higher dimensions the Hamiltonian is more complicated \cite{Paul,Mansfield10,Nair,Kogan,Green,Poljsak}. In order to solve the theory apart from solving the constraints corresponding to the symmetries of the theory you need to perform non-trivial calculations for determining the energy eigenvalues and the corresponding wave functionals. This does not diminish the contribution of the information provided from the constraints. In fact we shall see that in the electric representation the long range interactions are determined mainly by them through the implementation of particular constraints they impose on the field configurations. These, by their turn, incorporated in the Hamiltonian eigenvalue equation will give the particular interactions dictated from the theory. 

The advantage of working in the electric (E) representation \cite{Jiannis,Jack1,Jackiw1,Freedman,Freed1,Nair1,Jiannis10,Jiannis1,AJ} (rather than the vector potential (A) representation \cite{Paul,Rossi,Kerman,Jackiw2,Diakonov}) is that by separating the gauge degrees of freedom as the ones which rotate the $E$ vector the constraint becomes a multiplicative (non-differential) equation. Fixing the gauge within the above procedure by fixing the direction of $E$, a part of the gauge degrees of freedom are removed. The remaining ones, as we shall see for the case of the Yang-Mills and Abelian theories, are the rotations around the direction of the electric vector, $E$. Integrating these angles out constrains the field configurations via the generation of delta functionals. As a result {\it the quantum constraints are reduced to classical ones}, which are imposed as the arguments of these delta functionals. Still the procedure here is in some aspects simpler than the classical case: it is not necessary to solve the equations constraining the field configurations, but rather they can be used as definition equations for the divergence of the field, $\p_iE^i$, and can be substituted in the Hamiltonian when acting on the wave functional. In this way the terms in the Hamiltonian eigenvalue equation are reduced significantly.

To understand the above procedure, first we shall rigorously study the Abelian theory in the E representation of the field within the Schr\"odinger representation including static sources. There, the usual results of the vacuum energy and the Coulomb potential between the sources are derived \cite{Hatfield}. In addition the Fourier transformation of the E representation vacuum wave functional is performed, and the corresponding one in the A representation is derived where the sources appear naturally to be ``dressed'' \cite{Zarembo0,Lavelle}. Then we turn to the Yang-Mills theory with gauge group $SU(N)$, where by proceeding in the same way as in the Abelian theory the Gauss law constraint is solved, which imposes Abelian like constraints to the $N-1$ Cartan components of the electric field. The implications for the generation of the Coulomb and confining interactions are presented without solving the Hamiltonian eigenvalue problem \cite{Jiannis10}. 

For an introductory study of \Sc representation see \cite{Sym1,lusch,Sym2,Sym3,Jack}.

\section{Abelian Theory} 

As a first step in order to understand the E representation we should solve the Abelian theory, where we incorporate only static sources rather than fermionic fields \cite{Lavelle}. In the E representation the field variables are $E^i(x)$, while $A^i(x)=i\de /\de E^i(x)$. For the wave functional of the theory, $\Psi[E]$, the Gauss law with two sources of opposite unite charges is
\be
\p_iE^i(x) \Psi[E]= \Big(\de(x-x_0)-\de(x-x_1)\Big)\Psi[E] \,\, .
\label{ga1}
\ee
As the operator $\p_iE^i(x)$ acts multiplicatively on the wave functional, equation (\ref{ga1}) can only be solved by imposing on $\Psi[E]$ to vanish everywhere else accept when the field configuration, $E$, satisfies
\be
\p_iE^i(x) = \de(x-x_0)-\de(x-x_1)\,\, ,
\label{ga2}
\ee
which is the classical Gauss law of the theory. Hence the wave functional can be taken to have the form $\Psi[E]=\prod_x \de\Big(\p_iE^i(x) - \de(x-x_0)+\de(x-x_1)\Big)\Phi[E]$, where $\Phi[E]$ is a gauge invariant functional. This is the proper way to impose such a constraint in quantum field theory, where $E$ should take all possible configurations. 

There is another way to approach the Gauss law. Equation (\ref{ga1}) determines that under a gauge transformation $\Psi[E]$ should transform in the following way
\bq
\Psi[E] \rightarrow \Psi[E^U]=&&e^{\int d^3x E^i \p_i U^{-1} U} U(x_1)\Psi[E]U^{-1}(x_0)=
\no \no
&&
e^{-i \int d^3 x E^i(x) \p_i \th(x)-i\th(x_0) +i \th(x_1)} \Psi[E]
\label{trans1}
\eq
where $U(\th)\equiv \exp i \th $ is a $U(1)$ gauge group element. Note that $E^U=UEU^{-1}=E$, i.e. in the Abelian theory the electric vector, $E$, is gauge invariant as it does not change under a $\th$ rotation. So the wave functional $\Psi[E]$ should not get transformed under $U$. The phase factor in the transformation is produced from the phase factor $\exp -i\int E^iA^i$ when the Fourier transformation of the gauge invariant functional $\Psi[A]$ is taken (for more details see \cite{Jackiw}). This phase factor should be combined with the group elements $U(x_1)$ and $U^{-1}(x_0)$ and cancel out, which happens when the condition (\ref{ga2}) is imposed on the field configurations. Hence, the symmetry of the theory straightforwardly imposes the above constraint. Also, from (\ref{trans1}) we see that the solution of (\ref{ga1}) can be expressed as the group integration over $U(1)$:\footnote{This integration over the gauge degrees of freedom is similar to the procedure followed in \cite{Zarembo0}.}
\bq
\Psi[E]=&&\int \D u e^{-\int d^3x E^i \p_i u u^{-1}} u(x_1)u^{-1}(x_0) \Phi[E]=
\no \no
&&
\int \D \th e^{-i \int d^3 x E^i(x) \p_i \th(x)-i\th(x_0) +i \th(x_1)} \Phi[E] =
\no \no
&&
\prod_x \de \Big( \p_i E^i(x) -\de^{(3)}(x-x_0) + \de^{(3)} (x-x_1) \Big) \Phi[E] \,\, .
\label{fun0}
\eq
That is the integration over the gauge degree of freedom $\th$ gives the delta functional condition. The functional integration $\D \th$ is along the range $[0, 2 \pi]$, but it can be extended to $[- \infty, + \infty]$ by assuming that the charges of the theory are quantized, which makes the exponential periodic in $2 \pi$, as it should be \footnote{A requirement, which would be enough to impose the constraints by itself.}. Though, there is an overall factor inserted from this extension, which makes the delta functional approach one rather than infinity when its argument goes to zero. For this reason we shall take the action of the functional differentiations on this delta functional to vanish.

$\Phi[E]$ will be determined from the following Hamiltonian eigenvalue equation
\be
H \Psi[E]\equiv{1 \over 2} \int d^3x \Big(g^2E^i(x)E^i(x)+{1 \over g^2} B^i(x)B^i(x)\Big) \Psi[E]=  \E \Psi[E]
\ee
where
\be
B^i(x)\equiv\ep^{ijk}\p_jA^k(x)=i \ep^{ijk}\p_j{\de \over \de E^i(x)} \,\, .
\ee
The Hamiltonian in the momentum representation of the spatial coordinates takes the form
\be
H={g^2 \over 2} \int {d^3 p \over (2 \pi)^3} E^i(p) E^i(-p)-{1 \over 2 g^2} \ep^{ijk} \ep^{i \bar j \bar k} \int d^3p (2 \pi)^3 p_j p_{\bar j} {\de \over \de E^k(p)} {\de \over \de E^{\bar k} (-p)} \,\, .
\ee
Here the Fourier transformations are defined as 
\be
E^i(x)=\int {d^3p \over (2 \pi)^3} e^{i \vec{p} \cdot \vec{x}} E^i(p) \Sp \text{and} \Sp {\de \over \de E^i(x)}=\int d^3p (2 \pi)^3 e^{i \vec{p} \cdot \vec{x}} {\de \over \de E^i(p)} \,\, .
\ee
We can proceed in analogy with the A representation where the vacuum wave functional is exponentially quadratic in the magnetic field by defining a new ``magnetic'' field composed from the electric (multiplicative now) variable, $E$,  as
\be
\tilde B ^i(x)\equiv i \ep^{ijk}\p_jE^k(x) \,\, .
\ee
Hence an ansatz for the unknown functional (in the momentum space) is
\bq
&&
\Phi[E]=\exp \int {d^3k \over (2 \pi)^3} h(k)\tilde B^i(p) \tilde B^i(-p) =
\nonumber\\ \nonumber\\
&&
\exp -\int {d^3k \over (2 \pi)^3} h(k)\Big(k^2 E^i(k)E^i(-k)-k^iE^i(k) k^jE^j(-k)\Big)
\label{fun11}
\eq
where we have used the relation $\ep^{ijk}\ep^{i \bar j \bar k}=\de ^{j \bar j} \de^{k \bar k} -\de ^{j \bar k} \de^{k \bar j}$. The function $h(p)$ is unknown. The full wave functional, $\Psi[E]$, includes the delta functional mentioned before, while its condition can be now  applied on the exponential to simplify it by substituting the second term in the exponent with the delta functions denoting the positions of the sources. But on this exponential will act functional differentiations so they will give zeros when they see these delta functions; so we can neglect them without any loss of information. As a check we could keep in the exponential all the terms as in (\ref{fun11}) deriving at the end exactly the same result obtained from the simplified functional \footnote{A somewhat expected result as $k^iE^i(k)$ is longitudinal and the Hamiltonian should be ``transparent'' to these terms.}. From now on we shall disregard these terms as we do not want the delta functionals to act on $\Phi[E]$, but rather on the eigenvalues of the operators we are interested into.

Applying the Hamiltonian operator on $\Psi[E]$ we obtain
\bq
&&
H \Psi[E]=\Big[ 2\de^{(3)}(0)\int d^3p \,\,p^4 h(p) + {g^2 \over 2}\int {d^3p \over (2 \pi)^3} E^i(p)E^i(-p) -
\no \no
&&
{1 \over 2 g^2} \int {d^3p \over (2 \pi)^3} 4p^4 \,\, h^2(p) \Big( p^2 E^i(p)E^i(-p)- p^iE^i(p) p^jE^j(-p) \Big) \Big]\Psi[E]
\eq
Obviously the third term in the RHS can be made to cancel the second by choosing, $h(p)=g^2/(2|p|^3)$. But there is still dependence of the eigenvalue, $\E$, on the field from the fourth term! At this point the Gauss law comes to rescue with the delta functional constraints, which impose the condition (\ref{ga2}). Hence, we derive
\bq
&&
\E= \de^{(3)}(0)\int d^3p \,\,|p| + {g^2 \over 2}\int {d^3p \over (2 \pi)^3} {1 \over p^2} p^iE^i(p) p^jE^j(-p)=
\no \no
&&
\de^{(3)}(0)\int d^3p \,\,|p|+{g^2 \over 2}\int d^3x d^3y \,\, \int {d^3p \over (2 \pi)^3} {1 \over p^2} e^{-i \vec{p} \cdot (\vec{x} - \vec{y})}  \,\, \p_iE^i(x) \p_jE^j(y)=
\no \no
&&
\de^{(3)}(0)\int d^3p \,\,|p|+
\no \no
&&
\Sp {g^2 \over 2} \int d^3x d^3y \,\, {1 \over 4 \pi} {1 \over |x-y|} \Big( \de^{(3)}(x-x_0)-\de^{(3)}(x-x_1) \Big)\Big(\de^{(3)}(y-x_0)-\de^{(3)}(y-x_1) \Big) \Rightarrow
\no \no
&&
\Sp \Sp \Sp\E=\de^{(3)}(0)\int d^3p \,\,|p|-g^2{1 \over 4 \pi} {1 \over |x_1-x_0|} +V_{s.e.}^{0,1}
\eq
where $V_{s.e.}^{0,1}$ is the self energy of the sources. This is the well known result. The energy of the vacuum as well as the attractive Coulomb potential between the two sources is also easily derived in the A representation. 

We see that $\Psi[E]$ is exponentially quadratic in the field so it is possible to perform its Fourier transformation. For this we rewrite the delta functional as
\bq
&&
\prod_x \de \Big( \p_i E^i(x) -\de^{(3)}(x-x_0)+\de^{(3)}(x-x_1) \Big)=
\no \no
&&
\int \D \th \exp i \int { d^3 k \over (2 \pi)^3} \Big(i k_i E^i(k) -e^{-i \vec{k} \cdot \vec{x}_0}+e^{-i \vec{k} \cdot \vec{x}_1} \Big) \th(-k) \,\, .
\eq
Taking the Fourier transformation of $\Psi[E]$, we derive
\bq
&&
\Psi[A] \equiv \int \D E \exp \Big\{ -i \int {d^3p \over (2 \pi)^3}E^i(k) A^i(-k) \Big\} \Psi[E] =
\no \no
&&
\int \D E \D \th \exp \int {d^3 k \over (2 \pi)^3} \Big[ -{g^2 \over 2} {1 \over |k|} E^i (k) E^i (-k) - E^i(k)(iA^i (-k)+k^i \th(-k)) 
\no \no
&&
\Sp \Sp \Sp \Sp \Sp -i(e^{-i \vec{k} \cdot \vec{x}_0}-e^{-i \vec{k} \cdot \vec{x}_1} )\th(-k)\Big]=
\no \no 
&&
\int \D \th \exp -{1 \over 2 g^2} \int {d^3 k \over (2 \pi)^3} \Big[|k|A^i(k)A^i(-k)-2i|k|k^iA^i (k)\th(-k) +|k|^3\th(k) \th(-k)+
\no \no
&&
\Sp \Sp \Sp i2g^2 (e^{-i\vec{k}\cdot\vec{x}_0}-e^{-i\vec{k}\cdot\vec{x}_1})\th(-k) \Big]\Rightarrow
\no \no 
&&
\Psi[A]=\exp \Big\{ -{1 \over (2\pi)^2 g^2} \int d^3x d^3y {B^i(x)B^i(y) \over |x-y|^2} -i \La(x_0)+i \La(x_1) \Big\}\,\, ,
\eq
where $B^i(x)$ is the magnetic field constructed out of the multiplicative now $A^i(x)$ (see \cite{Hatfield}), while $\p_i\La(x)$ is the longitudinal component of $A^i(x)$ with
\be
\La(x)=-{1 \over 4 \pi}\int d^3y {\p_i A^i(y) \over |x-y|} \,\, .
\ee
The magnetic term $B^i(x)B^i(y)$ can be also written with respect to only transverse components of $A$. The exponential of $\La(x)$ as presented in the wave functional, can be combined with the fermionic sources denoted by operators multiplied on $\Psi[A]$ making them ``dressed'', that is gauge invariant! The dressing of the sources is necessary to give the proper asymptotic behavior to the fermions \cite{Lavelle}. However, performing the dressing with a Wilson line of the full $A$ vector rather than only its longitudinal component, would result to infinite terms in the potential as explained in \cite{papu}. That the delta functional evokes only the longitudinal component of $A$ can be seen by Fourier transforming the delta functional $\prod_x \de(\p_iE^i)$ alone
\be
\int \D E \D \th \exp -i \int {d^3 k \over (2 \pi)^3} E^i(k)\Big(A^i (-k)-ik^i \th(-k)\Big) = \int \D \th \de \Big(A^i (x)-\p_i \th(x)\Big)
\ee
which picks up {\it only} the longitudinal components of the vector potential for the dressing of the sources.

In addition, the simplification of the functional in the E representation due to the delta functionals is not possible in the A representation as there isn't such a condition there (the Gauss law is implied via the gauge invariance and $\La(x)$ terms for the sources). This simplification may be of small importance for the Abelian theory, but in the case of the Yang-Mills theory, where the terms needed to consider in the wave functional are many, a reduction of this form may be of significance. 

\section{Yang-Mills Theory}

Here, we shall study the more intriguing problem of Yang-Mills theory in the E representation \cite{Jack1,Jiannis10,Jiannis1}. The classical Hamiltonian is given by
\be
H=\int d^3 x \Big({g^2 \over 2} E^{ia}(x) E^{ia}(x) + {1 \over 2 g^2}B^{ia}(x)B^{ia}(x) \Big) \,\, ,
\ee
where $B^{ia}(x)=\ep^{ijk}(\p_jA^{ka}(x)+1/2f^{abc}A^{jb}(x)$ $A^{kc}(x))$, and it is accompanied by the Gauss law constraint
\be
\p_i E^{ia}(x)+if^{abc} A^{ib}(x) E^{ic}(x)=0 \,\, ,
\ee
where the index $a$ runs over the $N^2-1$ components of the $su(N)$ algebra. For quantizing in the electric Schr\"odinger representation the commutation relation
\be
[A^{ia}(x),E^{jb}(y)]=i \de^{ij} \de^{ab}\de^{(3)}(x-y)
\ee
is materialized by taking $E^{ia}$ diagonalized and $A^{ia}$ a differential operator, $A^{ia}(x)=i{\de \over \de E^{ia}(x)}$. The wave functional, $\Psi[E]$, has to satisfy the Hamiltonian eigenvalue equation $ H \Psi[E]=\E \Psi[E]$ and the Gauss law constraint
\be
\left(\p _iE^{ia}(x)-if^{abc}E^{ib}(x) {\de  \over \de E^{ic}(x)}\right)\Psi[E]=\Psi[E]T^a \de^{(3)}(x-x_0)-T^a\Psi[E] \de^{(3)}(x-x_1) \,\, .
\label{gauss1}
\ee
where a source is placed at $x_0$ and an antisource at $x_1$. Hence, the Gauss law enforces the following transformation property on $\Psi[E]$: 
\be
\Psi[E^U]=e^{{1 \over c}\tr \int E^i \p _iU^{-1} U} U(x_1)\Psi[E] U^{-1}(x_0) \,\, ,
\ee
similar to the Abelian case, but now $U\in SU(N)$ is a matrix. The solution of (\ref{gauss1}) can be given in terms of the following group integration
\be
\Psi[E]=\int \D u e^{-{1 \over c} \tr \int E^i \p_iu u ^{-1}} u(x_1)u^{-1}(x_0) \,\,\Phi[E]
\label{fun1}
\ee
The undetermined functional, $\Phi[E]$, is invariant under gauge transformations and will be calculated from the Hamiltonian eigenvalue equation. In contrast to the $A$ representation, where gauge invariant objects made out of $A$ are non-local, for example the trace of the Wilson loop, in the $E$ representation the gauge invariant objects are more elementary due to the vector character of the field. They are traces of products of the field $E$. 

We can separate the gauge degrees of freedom from the dynamical variables \cite{Jiannis}. As the gauge transformations rotate the $E$ vector in the $su(N)$ space we can consider one fixed direction, named $K^i$; then the general vector $E^i$ is obtained from it by an $SU(N)$ rotation, $g$, as $E^i(x)=g(x)K^i(x)g^{-1}(x)$, where $g$ transforms as $g(x) \rightarrow g^U(x)=U(x)g(x)$. Hence, the $SU(N)$ gauge symmetry of the electric field is clearly separated. $K^i$ can be used as the dynamical variable while $g$ can be fixed as its value should not have any physical relevancy. $K^{ia}$ has $3(N^2-1)$ degrees of freedom and $g$ has $N^2-1$. So the reparameterization of the electric field in terms of $K$ and $g$ needs additional constraints in order to become one-to-one. To define the appropriate constraints we need particular decomposition properties of $SU(N)$, which have some interest by themselves.

A general $SU(N)$ element, $g$, can be taken to satisfy
\be
{\cal J}^L_a g\equiv L_{ab}{\p \over \p \chi^b}g=-T^a g \,\,\,\,\, , \,\,\,\,\,\,\,\,\,\,\,\,\,\,\,\,\,\, {\cal J}^R_a g \equiv R_{ab}{\p \over \p \chi^b}g = g T^a \,\, ,
\label{diag}
\ee
for a special reparameterization, $g(\chi)$, with respect to the $N^2-1$ angles, $\chi_a$, and for some invertible matrices $L$ and $R$ \cite{Jiannis}. From the $N^2-1$ matrix hermitian generators, $T^a$, of the $SU(N)$ group, satisfying $\tr T^a T^b =c \de^{ab}$, it is convenient to let the Cartan elements to be the $N-1$ first ones, i.e. $(T^\ka)_{\al \bi}\equiv f^\ka(\al) \de_{\al \bi}$ for $\ka=1...N-1$. Along with this matrix representation of the generators the differential generators ${\cal J}_a^L$ or ${\cal J}_a^R$ can be organized similarly by diagonalizing the upper $(N-1)\times(N-1)$ block of $L$ and $R$. The first $N-1$ of the ${\cal J}$'s can be taken to be the differentiation with respect to a single angle, named $\phi^\ka$ for the ``left'' generators and $\bar \phi^\ka$ for the ``right'' ones. Then the first $N-1$ of the relations (\ref{diag}) become
\be
{\cal J}^L_\ka g\equiv -i{\p \over \p \phi^\ka}g=-T^\ka g \,\,\,\,\, , \,\,\,\,\,\,\,\,\,\,\,\,\,\,\,\,\,\, {\cal J}^R_\ka g \equiv -i{\p \over \p \bar \phi^\ka}g = g T^\ka \,\, ,\, \Sp \ka=1...N-1\,\,,
\ee
which can be solved to find the $\phi$ and $\bar \phi$ dependence of $g$, as 
\be
g(\chi)=h(\phi)\tilde g(\th) h(\bar \phi)
\ee
where $\theta$ are the remaining $(N-1)^2$ angles of $\chi$, and the $h$ elements belong to the Cartan subgroup, $H$, of $G=SU(N)$, while $\tilde g\in G/H$. The range of the angles is $\phi:\,\,[0,2 \pi]$, $\bar \phi:\,\,[0,2 \pi]$ and $\th: \,\,[0, \pi]$.

With the above diagonalization of the $SU(N)$ group it is possible to perform the group integration in relation (\ref{fun1}). The functional turns out to have only exponential dependence on $\bar \phi$, so the $\D \bar \phi$ integration produces a delta functional. (The $\bar \phi$ angles are similar to the $\th$ angle we met in the Abelian case. For the algebraic components of $K^i$ which belong in the Cartan subalgebra the group elements $h(\bar \phi)$, in relation $g K^i g^{-1}$, commute with it and cancel out! These $N-1$ components, $K^{i\ka}$, turn out to have Abelian-like behavior). The generated delta functionals make the $\D \theta$ and $\D \phi$ integrations easy to perform, resulting finally in \cite{Jiannis10,Jiannis1}
\bq
\Psi[E]&=&e^{- {1 \over c} \tr \int E^i \p_ig g^{-1}} \sum_\rho  g(x_1) P^\rho g^{-1} (x_0) 
\no \no
&&
\times \prod_{\ka=1}^{N-1} \prod_x \de \Big( \p_i K^{i\ka}(x)-f^\ka(\rho)\de^{(3)} (x-x_0) +f^\ka(\rho) \de^{(3)} (x-x_1) \Big) \,\, \Phi[E] \,\, ,
\label{fun2}
\eq
where $P^\rho \equiv \text{diag}(0...0,1,0...0)$, with $1$ in the $\rho$-th place, while its dimensionality depends on the representation of $SU(N)$ we have chosen. The first two terms of (\ref{fun2}) give the transformation properties of the wave functional required from the Gauss law. In the case where there are no sources, the constraints in the delta functionals are $ \p_i K^{i\ka}=0$ for $\ka=1...N-1$. It is natural then, to extend the constraints to $\p_i K^{ia}=0$ for $a=1...N^2-1$ so that spatial symmetry is restored and the decomposition $E^i=gK^ig^{-1}$ becomes one-to-one \cite{Zwanziger}. This symmetry is obviously broken when the sources are present. We shall keep the constraints for the rest of the components in mind without inserting them in a delta functional form.

A similar argument to the Abelian one for the generation of the constraints by conditions imposed from the phase factor produced from a gauge transformation cannot be applied here as a simple calculation shows that the phases produced from (\ref{fun2}) cancel identically. This is due to the existence of $\exp i\Omega =\exp - {1 \over c} \tr \int E^i \p_ig g^{-1}$ in the wave functional, which creates after a gauge transformation the proper phase factors for the cancelation. Note that $\Omega$ is not an analytic functional of $E$, as it depends also on $g$, and hence you cannot write $\Omega[E]$. Still the demand of periodicity of the angles $\bar \phi$ in $\Omega$ generates the above mentioned constraints on $\p_iK^{i \ka}$.

In order to determine $\Phi[E]$ the Hamiltonian has to be applied on $\Psi[E]$ and its eigenvalues should be sought. For this it will be necessary to fix the symmetry of the rotations with respect to $g \in SU(N)$, as these variables will cause extra divergences when the functional differentiations are taken \cite{Jack1,Freedman}. These divergences are not relevant for the dynamics of the theory because the physical variables have to be independent from the gauge rotations. Hence, we take $g= {\bf 1}$, and the wave functional becomes
\be
\Psi[K]=\sum_\rho  P^\rho \prod_{\ka=1}^{N-1} \prod_x \de \Big( \p_i K^{i\ka}(x)-f^\ka(\rho)\de^{(3)} (x-x_0) +f^\ka(\rho) \de^{(3)} (x-x_1) \Big) \,\, \Phi[K] \,\, .
\label{fun3}
\ee
The presence of the sources is still denoted with the delta functionals. The unknown part is the gauge invariant functional $\Phi[K]$, which will be determined from the \Sc equation. Note the similarity of this functional with the one in the Abelian case (\ref{fun0}). With the electric field fixed towards the $K^i$ direction the \Sc equation becomes
\be
H[K] \Psi[K] \equiv \int d^3 x \Big( { g^2 \over 2} K^{ia}(x)K^{ia}(x) +{1 \over 2 g^2} B^{ia}(x) B^{ia}(x)\Big)\Psi[K]= \E \Psi[K]
\label{ham1}
\ee
with
\be
B^{ia}(x)=i \ep ^{ijk} \Big( \p_j { \de \over \de K^{ka}(x)} + {i \over 2 } f^{abc} { \de \over \de K^{jb}(x)} { \de \over \de K^{kc}(x)} \Big) \,\, .
\label{magn1}
\ee
Here we shall not attempt to solve this equation. Though, it would be tempting to try an ansatz as in the Abelian case to be the exponential of $\int d^3x d^3y h(x,y) \tilde B^{ia}(x)\tilde B^{ia}(y)$ for $\tilde B$ being the modified magnetic field constructed out of $iK^{ia}(x)$ in the place of $A^{ia}(x)$. Merely, we shall be interested in determining the conditions for deriving the Coulomb and the confining potentials with the help of the delta functionals generated from the Gauss law or equivalently implied from the gauge symmetry of the theory.

Let us take the following functional
\bq
\Phi[K]&&=\exp \Big\{ \La_0 \int {d^3k \over (2 \pi)^3} {1 \over |k|} K^{ia}(k)K^{ia}(-k)+\La_1 \int {d^3k \over (2 \pi)^3} {1 \over |k|^2} K^{ia}(k)K^{ia}(-k)+...\Big\} \equiv
\no \no
&&
\equiv\exp W_{eff}[K] \,\, ,
\eq
where $W_{eff}$ is a proposed effective functional obtained after the renormalization of the wave functional is performed (see \cite{Mansfield10,MSP,Zarembo,PahOh,Jiannis10}). There are more terms necessary to fulfill the \Sc equation but we shall consider only these two as they will produce the potentials we are interested into. 

First note that the constant $\La_0$ is dimensionless, so very naturally this term could exist in the initial wave functional before renormalization. This term is the same as the Abelian one, and its existence in $\Psi[K]$ can be proven also perturbatively, having in the lowest order $\La_0=-g^2/2$, \cite{Mansfield10}. In the second term, the constant $\La_1$ is {\it dimensionful} ($\La_1 \sim [L]^{-1}$) and this should be a result of the regularization of the theory, which generates dynamically mass, $m$, as there is not any dimensionful constant in the initial theory. Hence, this constant, $\La_1$, can be written in terms of $m$. Its calculation would need to consider higher orders in perturbation theory.

For the following we shall also consider only the first term of the potential of the Hamiltonian (\ref{ham1})
\be
{1 \over 2 g^2} \int d^3 p (2 \pi)^3 p^i {\de \over \de K^{ia}(p)} p^j {\de \over \de K^{ja}(-p)}
\label{dyn1}
\ee
which as we have seen in the Abelian case gives the Coulomb potential. Due to its structure it generates longitudinal components of the electric field, which combined with the delta functional conditions produces the interactions. When it is applied on the first term it will give in the same way as before the desired Coulomb interaction between the static sources present also in the Yang-Mills theory. In fact, due to the matrix form of $\Psi[K]$ we have to take the trace of the expectation value of the Hamiltonian. Eigenstates of the Hamiltonian would be each term of the sum $\sum_\rho$ in (\ref{fun3}), while the state we are interested into is their superposition, which has the desired expectation value (explained in detail in \cite{AJ}). In addition a summation has to be performed over the $N+1$ directions the Cartan subalgebra can take in $su(N)$, which is a remnant of the group integration (\ref{fun1}). Hence, the Coulomb potential, $V_{Coul}$, coming from this term is given from
\be
V_{Coul}=g^2 {1 \over 4 \pi} C_2(N) {1 \over |x_1 - x_0|} + V_{s.e.}^{0,1}
\ee
where $C_2(N)$ is the quadratic Casimir invariant of $SU(N)$ and $V _{s.e.}^{0,1}$ is the self energy of the sources due to the Coulomb interaction. 

The application of (\ref{dyn1}) on the second term of $W_{eff}$ will give, up to an additive constant,
\bq
&&
{ 2 \La_1^2 \over g^2} \int d^3p {1 \over p^4} p_i K^{ia}(p)p_j K^{ja}(-p)=
\no \no
&&
{ 2 \La_1^2 \over g^2} \int d^3x d^3 y \int {d^3p \over (2 \pi)^3} {1 \over p^4} e^{-i \vec{p} \cdot (\vec{x} - \vec{y})} \p_iK^{ia}(x) \p_jK^{ja}(y)
\eq
We can easily calculate that
\be
\int{d^3p \over (2 \pi)^3} {1 \over p^4} e^{-i \vec{p} \cdot (\vec{x}-\vec{y})}={1 \over 4 \pi^2} \Big({\sin \ep \over \ep^2} + {\cos \ep \over \ep}- {\pi \over 2} \Big)|x-y|
\label{inf1}
\ee
where $\ep$ is going to zero. Hence, the term we are interested in becomes after averaging, tracing and multiplying by $N+1$ 
\be
V_{conf}={\La_1^2 \over 2 \pi g^2} C_2(N) |x_1 -x_0|+ \bar V _{s.e.}^{0,1} \,\, ,
\label{conf}
\ee
where $\bar V _{s.e.}^{0,1}$ is the self energy of the sources due to the confining interaction. It is worth noticing that this self energy (which is an infinite term) is produced from the evaluation of the integral in (\ref{inf1}) rather than the coincidence of the delta function arguments, which produces the self energies for the Coulomb interaction. This result reveals the dual character of the confining potential $\sim |x-y|$ with respect to the Coulomb one $\sim 1 / |x-y|$. For the second the infinities come from the identification of the points $x$ and $y$ while in the first case they come from the region around $p \rightarrow 0$ or equivalently $x \rightarrow \infty$.

\section{Discussion}

We have seen that Yang-Mills theory in the E representation, with the expression of the Gauss law with classical-like constraints obtains $N-1$ Abelian-like variables: the Cartan components of $K^{i}$. Moreover, the importance of the {\it longitudinal Cartan components} of the electric field is apparent for the propagation of the interaction between the sources. Their divergence, $\p_i K^{i\ka}$, is constrained with delta functionals in the same way as in the Abelian theory, resulting to their decouple from the Hamiltonian eigenvalue problem. This might bring the problem to a simpler form. Moreover, the classical constraints are just used as definition equations for $\p_i K^{i\ka}$, so you do not need to solve them and deal with classical configurations like monopoles, instantons and so on.

From this treatment we see that the quantum theory encloses in some sence a classical one. This could be connected with the success of the large $N$ study of gauge theories where the dominant classical behavior gives the basic characteristics of the theory. Even more there have been evidences for some theories that their results do not change much if you take $N$ to be quite small, signifying the dominance of the classical character in the full theory \cite{Dalley}.

We have seen how important is for the calculation of the confining potential to perform first the renormalization of the theory, where a dimensionful scale is generated. This will be used to change the conditions from the dimensionless factor of the Coulomb term, $1/|x-y|$, to the confining one, $|x-y|$, which needs to have a dimensionful factor ($\sim [L]^{-2}$) in order to be in dimensional agreement with $1/|x-y|$, as seen in (\ref{conf}).

In the A representation the dressing of the sources in the Yang-Mills theory is not a straightforward problem \cite{Lavelle}. In the E representation the insertion of the sources is denoted simply with their presence in the delta functionals. As we have seen for the Abelian case the dressing in the A representation can be obtained with a Fourier transformation of the functional $\Psi[E]$. But this is hard to perform in the Yang-Mills case where the logarithm of $\Phi[K]$ is of higher order than quadratic in $K$ and also undetermined! So this is an additional reason for working in the E representation for the study of fermion sources.

Of course there are disadvantages in the study of Yang-Mills theory in the E representation. In contrast to the kinetic term in the Hamiltonian, which is quadratic in $E$, the potential is up to quartic in $A$ so there are present quartic functional derivatives! This in the first sight complicates extremely the problem. Though, an approximate solution of the \Sc equation focusing on the zero propagating modes is possible with the consideration only of quadratic terms \cite{Jiannis10}. Its presence is enough to guaranty the existence of the main features of the theory like mass generation, Coulomb and confining potentials. Though, there is need of farther study of the dynamics of the theory and its behavior to achieve better quantitative results.

\end{document}